\documentclass[twoside,12pt]{article}
\usepackage{epsfig}
\usepackage{amsmath,amsfonts}

\newcommand{\be}{\begin{equation}}
\newcommand{\ee}{\end{equation}}
\newcommand{\bea}{\begin{eqnarray}}
\newcommand{\eea}{\end{eqnarray}}

\begin{document}

\title{A simple link of information entropy of  quantum and classical systems with Newtonian $r^{-2}$ dependence of Verlinde's  entropic force }

\author{C.P. Panos and Ch.C. Moustakidis \\
$^{}$ Department of Theoretical Physics, Aristotle University of
Thessaloniki, \\ 54124 Thessaloniki, Greece }

\maketitle

\begin{abstract}
It is shown that the entropic force formula $F_e=-\lambda\partial S/\partial A$ leads to a Newtonian $r^{-2}$ dependence. Here we employ the universal property of the information entropy $S=a+b\ln N$ ($N$ is the number of particles of a quantum system and $A$ is the area containing the system). This property was previously obtained for fermionic systems (atoms, atomic  clusters, nuclei and infinite Fermi systems  i.e. electron gas, liquid $^3$He and nuclear matter) and bosonic ones (correlated boson-atoms in a trap). A similar dependence of the entropic force has been derived very recently by Plastino et al with a Bose or Fermi gas entropy, inspired by Verlinde's conjecture~\cite{Verlide-11} that gravity is an emergent entropic force. Finally, we point out that our simple argument holds for classical systems as well.

%

%


\end{abstract}

The Shannon information entropy~\cite{Shannon-48,Birula-75} in position-space is:
\begin{equation}
S_r=-\int\rho({\bf r})\ln \rho({\bf r}) d{\bf r}
\label{S-r-1}
\end{equation}
and in  momentum-space:
\begin{equation}
S_k=-\int n({\bf k})\ln n({\bf k}) d{\bf k}
\label{S-k-1}
\end{equation}
where $\rho({\bf r})$, $ n({\bf k})$ are the position- and momentum-space density distributions respectively, which are normalized to one. We focus on the net information content, specifically the sum:
\begin{equation}
S=S_r+S_k
\label{S-sum}
\end{equation}
The merit of $S$ is that it does not depend on the unit of length in measuring $\rho({\bf r})$  and $ n({\bf k})$ i.e. the sum $S_r+S_k$ is invariant to uniform scaling of coordinates.  It is measured in bits if the base of the logarithm is $2$ and nats, if the logarithm is natural.  The entropic sum $S=S_r+S_k$ in conjugate spaces satisfies an entropic uncertainty relation (EUR), which for a three-dimensional system is
\begin{equation}
S=S_r+S_k \geq 3(1+\ln \pi)\simeq 6.434, \ (\hbar=1)
\label{EUR-1}
\end{equation}
established in~\cite{Birula-75}. The fact that $S_r+S_k$ is a dimensionless quantity, which contains information from both position and momentum-spaces, makes $S_r+S_k$ more appealing than $S_r$ or $S_k$ separately.
In \cite{Massen-98} we proposed a universal property for $S$ coming from  the density distributions of nucleons in nuclei, electrons in atoms and valence electrons in atomic clusters, as well as in~\cite{Massen-02} from trapped boson-alkali atoms in a correlated bosonic system:
\begin{equation}
S=a+b\ln N,
\label{S-N}
\end{equation}
where $N$ is the number of particles of a many-body quantum system, and $a$, $b$ constants, depending on the system under  consideration.
We have also found \cite{Moustakidis-05} that the same functional form holds for ideal infinite Fermi systems.

Specifically, the nuclear densities $\rho({\bf r})$  and $ n({\bf k})$  for several nuclei were obtained with Hartree-Fock calculations employing the Skyrme parametrization of the nuclear mean field. There exist various parametrizations of the Skyrme interaction, but they influence slightly the information entropies \cite{Lalazissis-98}. Thus we used the SKIII interaction \cite{Dover-72}. Finally, we fitted the form $S=a+b\ln N$ to the values obtained with SKIII interaction and found:
\begin{equation}
S=S_r+S_k\simeq 5.319+0.86 \ln N
\label{S-N-2}
\end{equation}
The fit is in reasonable good agreement with its H.O. counterpart \cite{Panos-97}.

In atomic (metallic) clusters the effective radial electronic potential was obtained by Ekardt \cite{Ekardt-84} in his spherical-jellium-background model study of the self-consistent charge density and the self-consistent effective one particle potential, using the local density approximation. Ekardt's potentials for neutral sodium clusters were parametrized in \cite{Nishioka-99}  by a Wood-Saxon potential of the form
\begin{equation}
V_{\rm WS}(r)=-\frac{V_0}{1+{\rm exp}[\frac{r-R}{a}]}
\label{WS-1}
\end{equation}
where: $V_0=6 \ eV$,  $R=r_0N^{1/3}$, $r_0=2.25 \AA$  and $a=0.74 \AA$. Details of the parametrization  of Ekardt's potentials can be seen in Ref.~\cite{Kotsos-97}. In \cite{Massen-98} we solved numerically the Schr\"{o}dinger equation for atomic clusters for several values of the number of valence electrons in the potential (\ref{WS-1}). Thus we found the wave functions of the single-particle states in configuration space and by Fourier transform the corresponding ones in momentum space. Employing the above wave functions, we calculated the electron density  $\rho({\bf r})$ in position space and $ n({\bf k})$ in momentum space, which were inserted in relations (\ref{S-r-1}), (\ref{S-k-1}). Hence we found the values of $S_r$, $S_k$. Then we fitted the form $S=a+b\ln N$ to these values and obtained the expression:
\begin{equation}
S=S_r+S_k\simeq 5.695+0.907 \ln N
\label{S-N-2}
\end{equation}
Similar forms exist in the literature for $N$ electrons in atoms \cite{Gadre-84,Gadre-85}
\begin{eqnarray}
&&S\simeq 6.65+\ln N \ (\text{Thomas-Fermi calculations})\nonumber\\
&&S\simeq 6.257+1.007\ln N \ (\text{Hartree-Fock})
\label{S-N-3}
\end{eqnarray}
In  Appendix I we review for completeness the justification of $S=a+b\ln N$ for ideal infinite Fermi systems. See also Ref.~\cite{Moustakidis-05}. In Appendix II we also derive $S=a+b\ln N \ (b=1)$ for a uniform classical system of $N$ particles.

The question naturally arises whether the same functional form for $S$ holds for bosonic systems. In fact, in Ref.~\cite{Massen-02} density distributions   $\rho({\bf r})$  and $ n({\bf k})$ for bosons were calculated by solving numerically the Gross-Pitaevski equation:
\begin{equation}
\left[-\frac{\hbar^2}{2m}\nabla^2+\frac{1}{2} m\omega^2r^2+ N\frac{4\pi\hbar^2\alpha}{m}|\psi({\bf r})|^2   \right]\psi({\bf r})=\mu\psi({\bf r})
\label{G-P-1}
\end{equation}
where $N$ is the number of atoms, $\alpha$  the scattering length of the interaction and $\mu$ the chemical potential. For a system of non-interacting bosons in an isotropic harmonic trap the condensate has a Gaussian form of average width $b$ ($b=\left(\hbar/m\omega\right)^{1/2}$).  If the atoms are interacting, the shape of the condensate can be changed considerably with respect to the Gaussian. The ground-state properties of the  condensate for weakly interacting atoms are explained quite successfully by the non-linear equation (\ref{G-P-1}). We solved it numerically (\cite{Massen-02}) for trapped boson-alkali atoms ($^{87}$Rb) systems with parameter $b=12180 \AA$ (angular frequency $\omega/\pi=77.78$Hz) and scattering length $\alpha=52.9 \AA$. In this case the effective atomic size is small compared both to the trap size and to the distance ensuring the diluteness  of the gas. The calculated  $\rho({\bf r})$  and $ n({\bf k})$  were inserted into (\ref{S-r-1}), (\ref{S-k-1}) to find the values $S_r$, $S_k$, $S$ as functions of the number of bosons $ N$. The fitting procedure gave:
\begin{equation}
S\simeq 6.033+0.068\ln N
\label{S-G-P-1}
\end{equation}
for $5\times 10^2 < N < 10^6$. We concluded in Ref.~\cite{Massen-02}  that a similar functional form holds approximately for $S$ as function of the number of particles $N$ for fermionic and bosonic systems (correlated atoms in a trap). It is remarkable that those properties hold for systems of different sizes i.e. ranging from the order of fermis ($10^{-13}$cm) in nuclei to $10^4 \AA$ ($10^{-4}$cm) for bosonic systems.

Now we use the entropic force formula~\cite{Plastino-18a,Plastino-18b}
\begin{equation}
F_e=-\lambda\frac{\partial S}{\partial A}
\label{Forc-1}
\end{equation}
$\lambda$ is a positive constant, $A=4\pi r^2$ is the area of the surface containing the system and $r$ its radius, i.e. $r=r_0N^{1/3}$.

We calculate $\frac{\partial S}{\partial A}$, using $S=a+b\ln N$. A combination of $A=4\pi r^2$ and $r=r_0 N^{1/3}$ gives
\begin{equation}
N(A)=\frac{1}{(4\pi r_0^2)^{3/2}}A^{3/2}
\label{N-2}
\end{equation}
Hence
\[\frac{dN(A)}{d A}=\frac{3}{2}\frac{1}{(4\pi r_0^2)^{3/2}}A^{1/2}  \]
Then
\[\frac{\partial S}{\partial A}=\frac{d S}{d  N}\frac{d N(A)}{d A}=b\frac{1}{N(A)}\frac{d N(A)}{d A}=\frac{3b}{8\pi}\frac{1}{r^2}  \]
Finally
\begin{equation}
F_e=-\frac{3b\lambda}{8\pi}\frac{1}{r^2}
\label{Forc-2}
\end{equation}
Choosing
\[\lambda=GMm\frac{8\pi}{3b}  \]
we obtain
\begin{equation}
F_e=-\frac{GMm}{r^2}
\label{Forc-3}
\end{equation}
The physical meaning for the choice of $\lambda$ is in accordance with  other authors that formula (\ref{Forc-3}) gives the entropic force equivalent to the force of gravity between a mass $M$ located at the center of the system under consideration and $m$ which lies on the periphery at distance $r$ from the center.

Relation $R=r_0 N^{1/3}$ is known to hold for nuclei and atomic clusters under the assumption of a constant density of nucleons and electrons respectively.
However, if we employ a more general form $R=r_0 N^{\alpha}$, ($\alpha>0$), we obtain the same $r^{-2}$ dependence i.e. $F_e=(-b\lambda/8\pi\alpha)/r^{2}$.  In fact, this is the case for our bosonic system (non-constant density), where we obtain for the rms radius $r=0.427 N^{0.191}$, which leads to the $r^{-2}$ dependence as well. There is one exception, relation $R=r_0 N^{\alpha}$ does not hold for electrons
in atoms, where the radius of the system oscillates as function of the atomic number, due to shell effects. Consequently our procedure can not
be applied to atoms.

It is noted that the form $S=a+b\ln N$ comes from considering a density normalized to one for a $N$-particle  system. This choice seems rather justified, because it guarantees $S>0$, since the physical meaning of the information entropy is that it represents the information content (number of bits or nats) of the system. One can easily transform to a density normalized to the number  of particles $N$, obtaining the form $\tilde{S}=\tilde{a}N+\tilde{b}N\ln N$ ($\tilde{a}=a$, $\tilde{b}=b-2$), but sometimes $\tilde{S}_r$ or $\tilde{S}_k$ or $\tilde{S}$ take negative values, a result  counter-intuitive for a quantity that measures information. Anyway, we check that our recipe $F_e=-\lambda\frac{\partial \tilde{S}}{\partial A}$  to  obtain $r^{-2}$ dependence does not work for $\tilde{S}$, due to the extra terms in $\tilde{S}$, which, however can be removed by changing the normalization of the density.

A transformation from $S$ to $\tilde{S}$ and vice versa is straightforward~\cite{Massen-01}. Let $\rho({\bf r})$ be the density distribution normalized to one and $\tilde{\rho}(r)=N\rho(r)$, the corresponding one normalized to $N$. Then
\begin{eqnarray}
S_r[\text{norm}=N]=\tilde{S}_r&=&-\int\tilde{\rho}({\bf r})\ln \tilde{\rho}({\bf r}) d{\bf r}\nonumber\\
&=&-\int N\rho({\bf r})(\ln N+\ln \rho({\bf r}))d{\bf r}=-N\ln N+NS_r
\label{Sr-norm-1}
\end{eqnarray}
Similarly
\begin{equation}
S_k[\text{norm}=N]=\tilde{S}_k=-N\ln N+NS_k
\label{Sk-norm-1}
\end{equation}
Inverting ~(\ref{Sr-norm-1}), (\ref{Sk-norm-1}) we find
\begin{equation}
S_r[\text{norm}=1]=S_r=\frac{\tilde{S}_r}{N}+\ln N
\label{Sr-norm-2}
\end{equation}
\begin{equation}
S_k[\text{norm}=1]=S_k=\frac{\tilde{S}_k}{N}+\ln N
\label{Sk-norm-2}
\end{equation}
Combining equations ~(\ref{Sr-norm-2}), (\ref{Sk-norm-2})  we have for the entropy sum $S=S_r+S_k$ and $\tilde{S}=\tilde{S}_r+\tilde{S}_k$
\begin{equation}
S=\frac{\tilde{S}}{N}+2\ln N
\label{S-Stil-1}
\end{equation}
and inverting
\begin{equation}
\tilde{S}=NS-2N\ln N
\label{S-Stild-2}
\end{equation}
The above transforms are useful for the following discussion. It will turn out that our result $F_e \sim r^{-2}$ does not depend on the specific value of $a$, but more importantly the normalization to one and $b>0$ are essential.

Now we discuss a concrete example from the literature. Gadre et al.~\cite{Gadre-84,Gadre-85} obtained $\tilde{S}_r=-5.59N-2N\ln N$ analytically with a TF (Thomas-Fermi) electron density for atoms normalized to $(N=Z)$ (number of electrons). A similar relation was also found in~\cite{Grassi-98} under the same conditions: $\tilde{S}_r=5.499N-2N\ln N$. One can check that $\tilde{F_e}=-\lambda \partial \tilde{S}_r/\partial A$ does not provide an $r^{-2}$ dependence of the entropic force due to the extra terms compared with the simple $S=a+b\ln N$ for norm one, employed in the present work. We make another point. Even if we transform Gadre's $\tilde{S}_r$ to $S_r=-5.59-\ln N$ the minus sign in the logarithm $\ln N$ would not give the correct sign (attractive) to the $r^{-2}$ dependence using $F_e=-\lambda \partial \tilde{S}_r/\partial A$. It is seen that a pragmatic approach is to use densities normalized to 1, employ the entropic sum  and  be careful to have $b>0$. All conditions are fulfilled  in our derivation.

A final comment seems appropriate: In a recent publication~\cite{Plastino-18a,Plastino-18b} Plastino et al., inspired by Verlinde's approach that gravitation is an entropic force, examined an example in the quantum case i.e. the Bose gas entropy~\cite{Lemons-014}
\begin{equation}
S=Nk_{B}\left[\left(\frac{n}{N} \right)\ln \left(1+\frac{N}{n} \right)+ \ln \left(1+\frac{n}{ N} \right) \right]
\label{S-Plas-1}
\end{equation}
where
\[n=V\left(\frac{E}{N} \right)^{3/2}\left(\frac{4\pi e m}{3h^2}  \right)^{3/2}  \]
$e$ is the base of the natural logarithm and $N/n$ is the average occupation number per cell i.e. the occupancy.
Taking into account that the volume can be cast as
\[V=\frac{4}{3}\pi r^3  \]
they calculated the entropic force~\cite{Plastino-18a,Plastino-18b}
\begin{equation}
F_e=-\lambda\frac{\partial S}{\partial A}
\label{Forc-4}
\end{equation}
In the classical limit $N/n \ll 1$ (low occupancy)
they obtained
\begin{equation}
F_e=-\lambda\frac{3 N k_B}{8\pi r^2}
\label{Forc-5}
\end{equation}
which is indeed of the Newton appearance, so that  Verlinde's conjecture gets proper in the classical limit.

In an analogous work Plastino and Rocca~\cite{Plastino-018} starting from the Fermi gas entropy~\cite{Lemons-014}
\begin{equation}
S=Nk_B\left[\ln\left(\frac{n}{N}-1  \right)-\left(\frac{n}{N}  \right)\ln\left(1-\frac{n}{N}\right)  \right]
\label{S-Plas}
\end{equation}
found (\ref{Forc-5}), although some details of the derivation are different than the bosonic case, as expected. In both cases their result is $F_e=-\mu/r^2$, where $\mu=\lambda 3N k_B/8\pi$ depends on the specific value of $N$ of the system. In our case $S=a+b\ln N$, $a$ is eliminated by the derivative $\partial S/\partial A$, while $b$ is of the order $1$ for fermionic systems ($b=1$ for infinite Fermi system, $b=0.907$ for atomic clusters, $b=0.86$ for nuclei). Our bosonic case is different ($b\simeq 0.068$, a weaker dependence on $N$). In general we find $F_e=-\nu/r^2$, where $\nu=3b\lambda/8\pi$. Again the coefficient  $\nu$ depends on the system, this time via  $b$. Both teams obtain  the $r^{-2}$ distance dependence for the entropic force of quantum systems, but they are not able at present  to correlate explicitly  with the standard gravitation constant $G$. However, they may contribute to the ongoing discussion~\cite{Zhi-018}, whether an information-theoretic approach or a thermodynamic one is suitable to explain the origin of gravity as an entropic force caused by changes in the information associated with the position of material bodies~\cite{Verlide-11}. The link of $\tilde{S}_r$ to the correlation energy~\cite{Grassi-98}   and the interpretation of the entropy sum $S_r+S_k$ as a correlation measure~\cite{Guevara-03},   together with Collins conjecture~\cite{Esquivel-96}: {\it  the correlation entropy is proportional to the information entropy}, might be a way forward.  At present this does not seem applicable to the entropic force derivation.

The entropy sum can be seen in a more general context starting from a phase-space distribution which depends on both position and momentum, for example the Wigner function~\cite{Laguna-013}, whose marginals are the densities in position and momentum spaces. Thus the entropy sum can be considered as the Shannon entropy of a separable phase-space distribution, which is the product of the densities in position and momentum spaces. Hence the entropy sum can be interpreted as a measure of position-momentum correlations in quantum systems providing a link to quantum  information theory for  research with continuous variables.

\section*{Appendix I}
Here, we review the derivation of $S=a+b\ln N$, obtained previously~\cite{Moustakidis-05} for ideal infinite Fermi systems.
The starting point for the description of the density distribution $\rho({\bf r})$ both in infinite and finite quantum systems is
the one-body density matrix (OBDM). The OBDM is defined as
\begin{equation}
\rho({\bf r}_1,{\bf r}_1')=\int \Psi^*({\bf r}_1,{\bf r}_2,\cdots,{\bf r}_N)\Psi({\bf r}_1',{\bf r}_2,\cdots,{\bf r}_N)d{\bf r}_2\cdots{\bf r}_N
\label{OBDM-1}
\end{equation}
The diagonal elements $\rho({\bf r}_1,{\bf r}_1)$ of the OBDM yield the local density distribution, which is just a constant $\rho$ in a uniform  infinite system. Homogeneity and isotropy of the system imply that $\rho({\bf r}_1,{\bf r}_1')=\rho(|{\bf r}_1-{\bf r}_1'|)\equiv \rho(r)$. In the case of a  noninteracting  Fermi system the associated OBDM becomes
\begin{equation}
\rho(r)=\rho l(k_F|{\bf r}_1-{\bf r}_1'|),
\label{rho-fer-1}
\end{equation}
where
\[l(x)=3x^{-3}(\sin x-x\cos x)  \]
and $\rho= N/V$ is the constant density of the uniform Fermi system.

The normalization $\int \rho_0 d{\bf r}=1$ leads to  the relation
\begin{equation}
\rho_0=\frac{1}{N V_0}=\frac{1}{N \frac{4}{3}\pi r_0^3}
\label{rfo-2}
\end{equation}
where the volume $V_0= \frac{4}{3}\pi r_0^3$ corresponds to the effective volume of the Fermi particle and $N$ is the number of fermions. The momentum distribution $n({\bf k})$ for fermions, having single-particle level degeneracy $\nu$, is defined by
\begin{equation}
n(k)=\nu^{-1}\int \rho(r) e^{i{\bf k}{\bf r}} d{\bf r}
\label{MD1}
\end{equation}
($\nu=2$ for electron gas and liquid $^3$He, and $\nu=4$ for nuclear matter).
The normalization $\int n(k) d{\bf k}=1$ leads to
\begin{eqnarray}
n(k)&=&\frac{1}{V_k}\left\{\begin{array}{ll}
\tilde{n}(k^-), \qquad k<k_F                \\
\\
\tilde{n}(k^+), \qquad k>k_F    \
                              \end{array}
                       \right.
\label{Basis-3}
\end{eqnarray}
where $V_k=\frac{4}{3}\pi k_F^3$. The Fermi wave number $k_F$ is connected with the constant density $\rho=N\rho_0=3/(4\pi r_0^3)$ as follows
\[k_F=\left(\frac{6\pi^2\rho}{\nu}  \right)^{1/3}=\left(\frac{9\pi}{2\nu r_0^3}  \right)^{1/3}.  \]
In the case of an ideal Fermi gas $n({\bf k})$ has the form
\begin{equation}
n_0(k)=\frac{1}{V_k}\theta(k_F-k)
\label{theta-1}
\end{equation}
The information entropy $S_r$ in coordinate-space (for density $\rho_0$ normalized to $1$) for a correlated or uncorrelated Fermi system is
\begin{equation}
S_r=-\int \rho_0\ln \rho_0 d{\bf r}=\ln V.
\label{S-ap-1}
\end{equation}
Considering that $V=NV_0$, $S_r$ becomes
\begin{equation}
S_r=\ln\frac{4}{3}\pi r_0^3+\ln N.
\label{S-ap-2}
\end{equation}
The information entropy $S_k$ in momentum space (for $n(k)$ normalized to $1$) is given by the relation
\begin{equation}
S_k=-\int n(k)\ln n(k) d{\bf k}
\label{S-ap-3}
\end{equation}
$S_k$ for an ideal Fermi gas, using Eq.~(\ref{theta-1}), becomes
\begin{equation}
S_k=\ln V_k=\ln \left(\frac{6\pi^2}{\nu r_0^3}  \right)
\label{S-ap-4}
\end{equation}
Combining  Eq.~(\ref{S-ap-2}) and (\ref{S-ap-4})  the information entropy sum $S=S_r+S_k$ for an uncorrelated infinite Fermi system becomes
\begin{equation}
S_0=S_r+S_k=\ln\left(\frac{8\pi^3}{\nu} \right)+\ln N
\label{S-ap-5}
\end{equation}
It turns out that the functional
\[S_0=a+b\ln  N  \]
for the entropy sum as a function of the number of particles N holds for an ideal infinite Fermi
system. Specifically for $\nu=2$ (electron gas and liquid $^3$He) $S=4.820+\ln N$, while for   $\nu=2$ (nuclear matter) $S=4.127+\ln N$.   The same function has been found in Ref.~\cite{Gadre-84,Gadre-85} for atoms and in Ref.~\cite{Massen-98} for nuclei and
atomic clusters.

\section*{Appendix II}
We consider a classical system of $N$ particles described by a spherical density distribution
\begin{eqnarray}
\rho(\tilde{r})&=&\left\{
\begin{array}{ll}
\rho_0, \quad 0<\tilde{r}<r     &          \\
\\
0, \qquad  \tilde{r}>r.  &  \
                              \end{array}
                       \right.
\label{rho-1}
\end{eqnarray}
normalized to $1$:
\[\int_0^r \rho(\tilde{r}) 4\pi \tilde{r}^2 d\tilde{r}=1  \]
which gives
\[\rho_0=\frac{3}{4\pi r^3}  \]
The information entropy of the distribution $\rho(\tilde{r})$ is defined as:
\begin{equation}
S=-\int \rho(\tilde{r})\ln\rho(\tilde{r}) 4\pi \tilde{r}^2 d\tilde{r}=-\int_0^r\rho_0\ln\rho_0 4\pi \tilde{r}^2 d\tilde{r}
\label{S-1}
\end{equation}
We obtain:
\begin{equation}
S=-\ln\left(\frac{3}{4\pi}\right)+3\ln r=1.432+3\ln r
\label{S-2}
\end{equation}
Using $r=r_0N^{1/3}$ which holds for a uniform $\rho(\tilde{r})$ (constant density) we find:
\begin{equation}
S=(1.432+3\ln r_0)+\ln N
\label{S-3}
\end{equation}
or:
\begin{equation}
S={\alpha}+\ln N \qquad (\alpha=1.432+3\ln r_0)
\label{S-4}
\end{equation}
It turns out that the specific value of $r_0$  is not relevant for our final conclusion.


\end{document}